# STEEP SPECTRUM SEYFERT GALAXIES

K.A. POUNDS
*X-ray Astronomy Group, Department of Physics and Astronomy, University of Leicester, Leicester, UK*

W.N. BRANDT
*Institute of Astronomy, University of Cambridge, Cambridge, UK*

**Abstract**

The realization that a substantial minority of Seyfert-type AGN exhibit unusually strong and steep soft X-ray spectra is reviewed, building on observations made during the *ROSAT* all-sky survey. Two individual cases are described in brief detail, both being identified initially in the small sample of Seyferts detected in the Wide Field Camera (WFC) survey. In one case the similarity in X-ray appearance to a high state Galactic Black Hole Candidate (GBHC) is noted, while the flare-like nature of the second example could represent a detection of the predicted stellar capture process. Extending the review to include the larger (and growing) sample of steep spectrum Seyfert-type galaxies being found with the *ROSAT* PSPC, we draw attention to the more common properties of the group, namely steep spectra, rapid variability, strong Fe II emission and identification with narrow-line Seyfert 1 galaxies.

## 1. Introduction

X-ray observations have presented a set of powerful diagnostics of Seyfert galaxies over the past decade (see Mushotzky et al. 1993 for references). First, a series of 'long looks' during the last months of *EXOSAT* produced several remarkable light curves of bright Seyfert galaxies, revealing for the first time that rapid, large-amplitude flux changes were a common property. Simple light travel time arguments showed the emission regions to be extremely small, providing strong circumstantial evidence for accretion onto a central black hole as the source of power. Broad-band X-ray spectroscopy provided the next revelation, with *Ginga* spectra showing the effects of re-processing in cold, dense matter ('reflection') of the hard power law characteristic of AGN in the 2–20 keV band. Then, hints in *Einstein* and *Ginga* data of complex low energy absorption features were dramatically confirmed with *ROSAT*, demonstrating that a substantial, but hitherto unsuspected, column of partially ionised ('warm') gas lay in the line of sight to several Seyfert nuclei.

A key spectral feature arising from the 'reflection' process is a strong K-fluorescence line of iron near 6.4 keV. Now, with the enhanced spectral resolution provided by the CCD cameras on *ASCA*, significant line broadening of this emission feature has been demonstrated (e.g. Tanaka et al. 1995). The preceding paper by Prof. Fabian has described how the observed line profiles yield evidence that the X-ray emission arises in a region of strong gravity, implicitly from an accretion disc extending very close to the event horizon of a massive black hole.

Thus, the past decade has provided observational data bearing directly on the nature of the compact energy source in Seyfert galaxies and on the form of matter in its immediate environment. While accretion is clearly the driver, the mechanism of the primary X-ray emission remains unclear. The basic power-law spectral form, with an underlying photon index (hereafter $\Gamma$) close to 1.9 in the 2–10 keV band for the majority of Seyferts, suggests a non-thermal process, a conclusion supported by both theoretical and observational indications that the thick accreting gas remains too cool for thermal emission to dominate in this 'hard' X-ray band (optically thin thermal Comptonization is, of course, possible). Indeed, even the 'soft excess' often seen as a steepening of Seyfert spectra below $\sim 1$ keV has been interpreted as arising from re-emission of the fraction of hard photons absorbed (rather than reflected) in the postulated disc. Evidence for this view arises from near-coincident variability in the hard and soft components, and from the soft excess typically having a luminosity significantly less than that of the hard power law.

Now, however, an increasing number of Seyfert 1 galaxies having unusually strong soft X-ray emission are being found in $ROSAT$ observations. This shows up, typically, as a steeply rising $ROSAT$ spectrum ($\Gamma > 3$, compared with the mean value of $\sim 2.3$ for 'normal' Seyferts at 0.1–2 keV), and it appears that as many as $\sim$10–15 per cent of Seyfert galaxies may be of this type. In this paper an account of this major sub-set of Seyferts is presented, based in part on the follow-up study of a small sample of sources identified in the extreme ultraviolet sky survey with the $ROSAT$ Wide Field Camera (WFC). Two particular sources are described in some detail, RE J 1034+393 for which we also now have $ASCA$ spectral data (and speculate that it represents the Seyfert analogue to the high state of a GBHC source) and RE J 1237+264, apparently first detected in a flaring state. We point out that the latter case fits the theoretical description of the burst of radiation that would be expected to follow the tidal disruption of a star just prior to capture by the putative black hole in the nucleus of the host galaxy. Our thesis is that the factor common to both soft X-ray sources (and the soft spectrum Seyferts as a whole) is most probably an unusually high accretion rate.

We note, finally, that parallel $ROSAT$ studies have now revealed over 30 Seyfert-type galaxies with best-fit $\Gamma > 3$ in the $ROSAT$ energy band, following a search based on a sample of narrow-line Seyfert 1 (NLS1) objects (e.g. BBF96; Brandt 1995; see the references in these papers to the optical literature on NLS1). Together with the $ROSAT$ all-sky survey Seyferts of Walter & Fink (1993), these new X-ray spectra highlight the absence of any steep spectrum objects also having permitted Balmer lines with FWHM $\gtrsim 3000$ km s$^{-1}$.

## 2. The ROSAT Wide Field Camera Seyferts

Five Seyfert galaxies were among the small sample of AGN detected in the WFC sky survey (Pounds et al. 1993; Pye et al. 1995). All were, understandably, in directions of low Galactic column density, and follow-up $ROSAT$ PSPC observations showed that four had unusually steep soft X-ray spectra. With hindsight, it is evident that this combination of properties was (just) sufficient to allow detection of these extragalactic sources in a band restricted (by the boron filter coating) to photons below 0.18 keV.

The unusually soft spectrum of one of the brightest WFC AGN, RE J 1034+393, was reported by Pounds (1994) and Puchnarewicz et al. (1995). The remarkable spectral softness in the $ROSAT$ band is illustrated in Figure 1, which compares the PSPC counts spectrum of RE J 1034+393 with that of the 'normal', bright Seyfert 1 galaxy NGC 5548 (both Seyferts have similar Galactic columns). Modelling the PSPC spectrum of RE J 1034+393 with the XSPEC software quantified this soft emission in terms of black bodies of $\sim 40$ and $\sim 100$ eV, with a 0.1–2 keV flux an order of magnitude higher than that of the (barely seen) hard power-law component (Pounds et al. 1995). Subsequent analysis of the PSPC spectrum of RE J 1034+393 from the $ROSAT$ all-sky survey observation, made in November 1990, revealed an essentially identical spectrum and flux level to the pointed observation one year later (H.H. Fink, private communication).

The WFC optical identification programme (Mason et al. 1995) included an observation of the field around RE J 1034+393 with the Isaac Newton Telescope Faint Object Spectrograph in May 1991. It revealed a Seyfert 1 galaxy, albeit one with unusually narrow permitted line widths (H$\beta$ FWHM $\sim 1500$ km s$^{-1}$; see Puchnarewicz et al. 1995 for a detailed discussion of the optical spectrum).

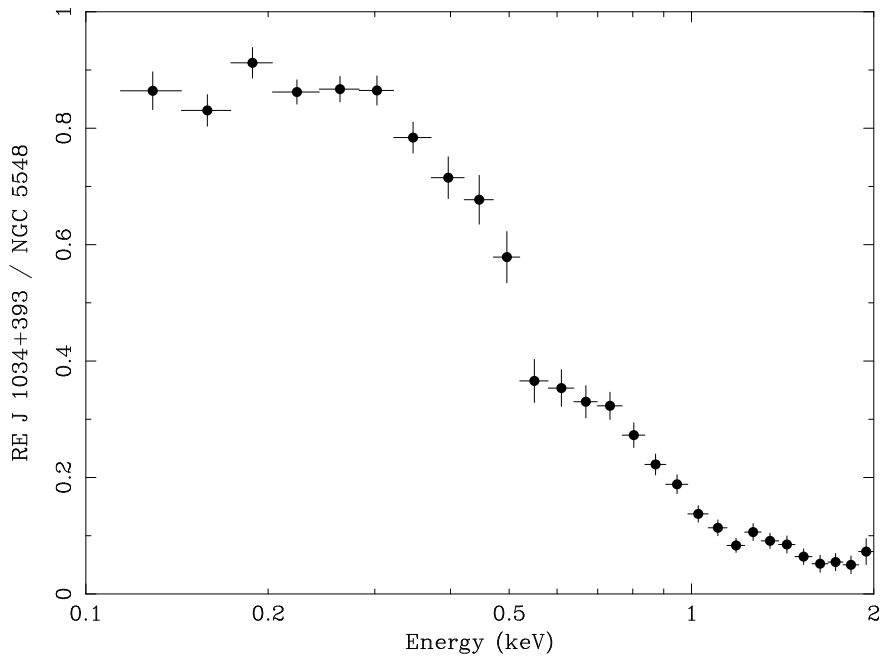

Fig. 1. Ratio of normalised pulse height spectra from PSPC observations of RE J 1034+393 and NGC 5548. Note the remarkable softness of RE J 1034+393. Both sources have similar Galactic column values.

Speculation that the strong soft X-ray emission indicated that RE J 1034+393 was in a high accretion rate state (Pounds 1994), analogous to the high states of GBHC, led to the prediction that Compton cooling by the intense soft photon flux would also steepen the harder power law above 2 keV. In order to test this analogy an *ASCA* observation was carried out in November 1994. This found RE J 1034+393 to still be in its high flux state, with a strong soft excess below $\sim 1$ keV (Figure 2). It also confirmed (Pounds et al. 1995) the prediction that the hard power-law component was indeed unusually steep, with a best-fit photon index of $\approx 2.6$. Interestingly, this index is consistent with the typical hard power-law slopes found in the high flux states of GBHC sources (see table 1 of Ebisawa et al. 1996 and references therein).

Subsequent analysis of PSPC data from the four other WFC Seyferts has shown three to have photon indices in the high range 2.8–4 in simple power-law fits over the 0.1–2 keV band. We draw particular attention here to one of these, RE J 1237+264 (discussed in more detail in Brandt et al. 1995; Grupe et al. 1995 and Brandt 1995), since it is also remarkable in its high level of variability. Indeed, only during the *ROSAT* all-sky survey observation in December 1990 was RE J 1237+264 in a high flux state (with a *ROSAT* band luminosity of $\sim 1 \times 10^{44}$ erg s$^{-1}$). Subsequent pointed observations in December 1991 and June–July 1992 showed it to have dimmed by a factor of over 70 (see Figure 3). Although less well constrained, the low-flux spectra appear about as soft as the survey spectrum, with $\Gamma \sim 4$. It is interesting to note that the retention of a soft spectrum during part of the transition to a low flux state is also found with GBHC sources. Miyamoto (1996) has described the hysteresis in the spectral evolution accompanying major flux level changes in the GBHC. We note, in this context, an intriguing report by Mannheim et al. (1996), who observed the transition of a steep spectrum *ROSAT* Seyfert, RX J 0134−42, to a 'normal' hard power law with little overall flux change, inviting the speculation that this observation caught the low flux transition from a soft to a hard spectrum. In the case of RE J 1237+264 the likelihood that the all-sky survey observation detected the source in a relatively brief high state, or 'flare', is strengthened on the basis of optical spectra taken over the period 1971–1995 (see Grupe et al. 1995; Brandt 1995 and references therein). Spectra taken in 1971, 1976, and annually from February 1992 to March 1995, show much fainter emission lines than in May 1991 shortly after the $ROSAT$ all-sky survey detection (Figure 4).

An exciting possibility as to the nature of the X-ray flare in RE J 1237+264 is that it could be the sighting of a stellar tidal disruption, as discussed by, for example, Hills (1975), Pravdo et al. (1981) and Rees (1988) and followed up initially using $ROSAT$ WFC data by Sembay & West (1993). Although the frequency and timescales of such phenomena are uncertain, a flare lasting of order a year lies within the range of prediction, while the expected high accretion rate is likely to lead to radiation peaking in the extreme ultraviolet band. In the context of such an explanation, it is interesting that the optical counterpart of RE J 1237+264, Zwicky 159.034 (a.k.a. IC 3599), exhibits some Seyfert properties in its 'quiescent' state. While at first sight finding evidence for stellar capture in a minority class of galaxies would appear surprising, this could be a consequence of a common trigger of both Seyfert activity and the likelihood of stellar capture (e.g. a galaxy interaction). Major inner accretion disc instabilities might also have caused the outburst of RE J 1237+264.

The remaining two steep spectrum WFC Seyferts, RE J 1337+242 (IRAS 13349+2438; see Brandt et al. 1996) and RE J 1442+352 (Mrk 478; see Gondhalekar et al. 1994), are also identified with NLS1, with the former showing additional evidence for line of sight absorption in an ionised, but dusty, medium. If we see this medium because we are viewing the X-ray core of RE J 1337+242 along the edge of a torus (see Brandt et al. 1996 and references therein for a discussion of why this might well be the case), this argues against the steep spectrum and other NLS1 properties being the consequence of viewing an accretion disc face-on (a conclusion supported more generally by the work of Boroson 1992 and Boroson & Green 1992). The relatively narrow permitted lines in NLS1 may instead be due to the effects of the intense flux of soft photons on the permitted line region (Guilbert et al. 1983; Bechtold et al. 1987; Brandt et al. 1994; Laor et al. 1994 and Laor et al., in preparation discuss this in more detail).

## 3. The NLS1 survey with ROSAT

Recognition that in the all-sky survey sample of Seyfert galaxies (Walter & Fink 1993), a sub-set having simple power-law fits with $\Gamma > 3$ were mostly NLS1 (Brandt et al. 1994), led BBF96 to search the $ROSAT$ pointed data archive for observations of a larger, optically-selected sample of NLS1. The result of that search was the finding that NLS1 as a class are generally steeper in the $ROSAT$ band than Seyfert 1s with broader permitted lines (see also Laor et al. 1994 who found this relation for quasars). Figure 5 plots the power-law slope and FWHM of the H$\beta$ line for these and some additional NLS1, together with the 'normal' Seyfert 1 galaxies in Walter & Fink (1993). The correlation of steep X-ray slope and low permitted line width is clear, as is the 'zone of avoidance' where Nature appears not to permit steep spectra and broad permitted lines to occur together.

A systematic ultrasoft NLS1 survey has also been performed using $ROSAT$ data by Grupe et al., in preparation. This survey has found many new NLS1. Several new $ROSAT$ NLS1 have also been discovered by, for example, Moran et al. (1996) and Greiner et al. (1996).

## 4. Other evidence of steep spectrum Seyfert galaxies

Although $ROSAT$ spectra have brought evidence of the steep spectrum Seyferts to widespread attention, a search of the *Einstein* data base for ultrasoft sources also led to the conclusion that a subset of AGN have unusually strong soft X-ray emission (Córdova et al.

1992). In addition, a substantial fraction of the *Einstein* sample were optically identified with NLS1 (Puchnarewicz et al. 1992).

Historically, it seems likely that the (at that time) uniquely soft Seyfert, H 1615+061, discovered in the *HEAO-1* survey in 1977 (Pravdo et al. 1981), and found several years later via *EXOSAT* observations to have 'reverted' to a normal hard power law (Piro et al. 1988), was an early example of the steep spectrum Seyferts. However, the available optical spectra of H 1615+061 interestingly appear to be those of a 'normal' Seyfert galaxy. White et al. (1984) also briefly discuss the early ultrasoft Seyferts (including NGC 4051, which satisfies the criteria for NLS1; see BBF96).

## 5. Discussion

A review of the general X-ray and optical properties of the steep spectrum Seyferts (e.g. BBF96) suggests that the two examples described in some detail in this paper are extreme rather than typical cases. While common properties are $\Gamma > 3$ and an optical NLS1 spectrum, many of the larger *ROSAT* sample show rapid X-ray variability (unlike RE J 1034+393; see BBF96; Otani et al. 1996 and Forster & Halpern 1996 for examples and references) and strong Fe II (unlike RE J 1034+393 or RE J 1237+264). A major shortfall of current observations is the scarcity of hard X-ray data on NLS1. It remains of great interest whether RE J 1034+393 is unique in having a steep power law extending into the 2–10 keV band. If this does turn out to be rare (as the prevalence of strong Fe II in many of the NLS1 sample may suggest), then the degree of Compton cooling implied in our explanation of the RE J 1034+393 spectrum may also be attained quite rarely. Whether or not RE J 1034+393 turns out to be a rare case of the Seyfert analogue to the high state of a GBHC, as suggested by Pounds et al. (1995), it seems most likely that the underlying and common feature of the steep spectrum Seyfert class is an unusually high accretion rate. If so, this offers a further important X-ray diagnostic of Seyferts, with the potential to study the observational consequences of a major difference in the primary driver of these accretion-powered sources, either across a sample of Seyferts or sometimes within the same object.

## 6. Acknowledgments

We acknowledge Th. Boller, C. Done, A. Fabian, H. Fink, J. Osborne and L. Puchnarewicz for help and useful discussions. WNB thanks the United States National Science Foundation for support.

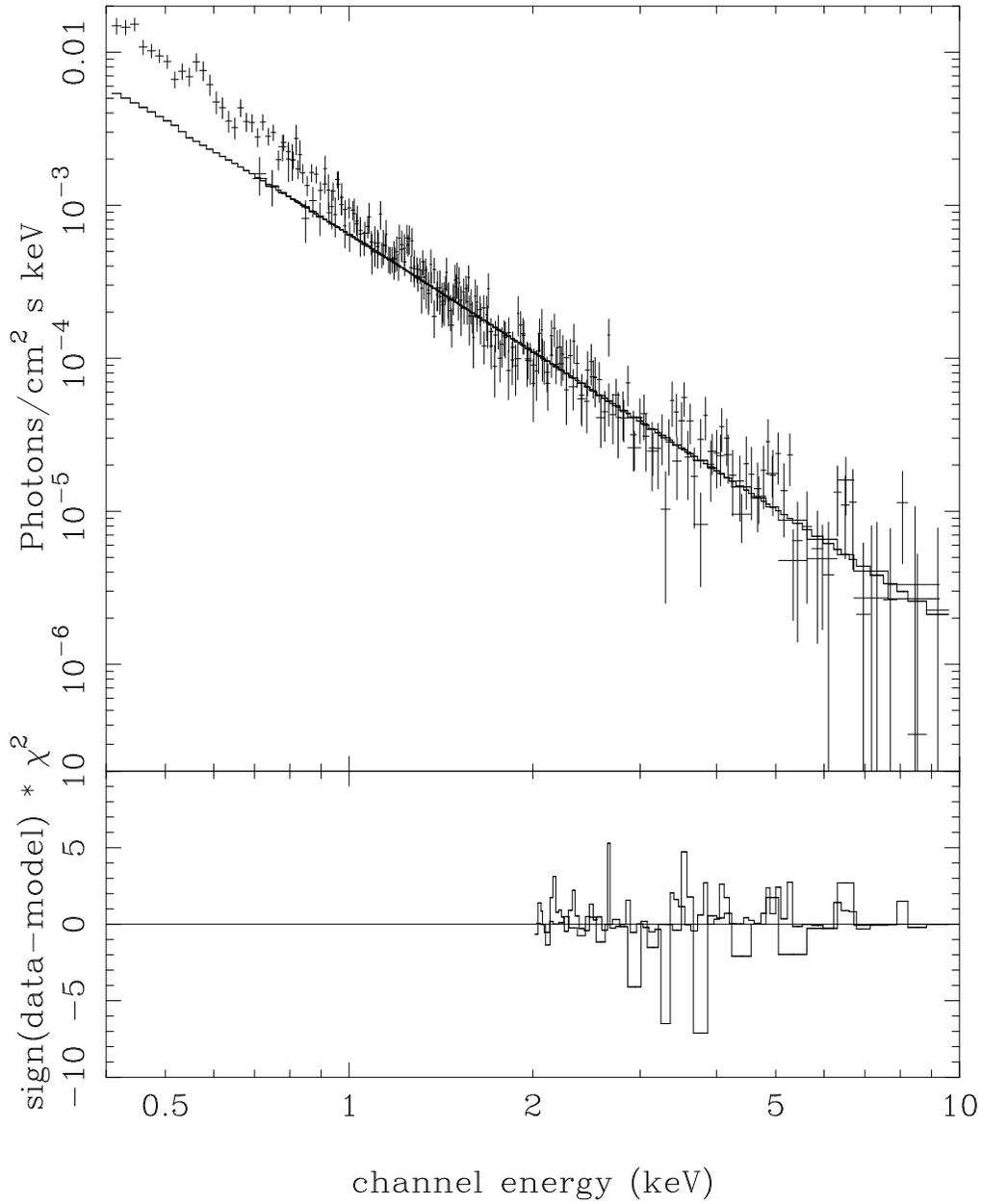

Fig. 2. *ASCA* spectrum of RE J 1034+393. Note the strong soft excess below ∼ 1 keV and the unusually steep hard power law. From Pounds et al. (1995).

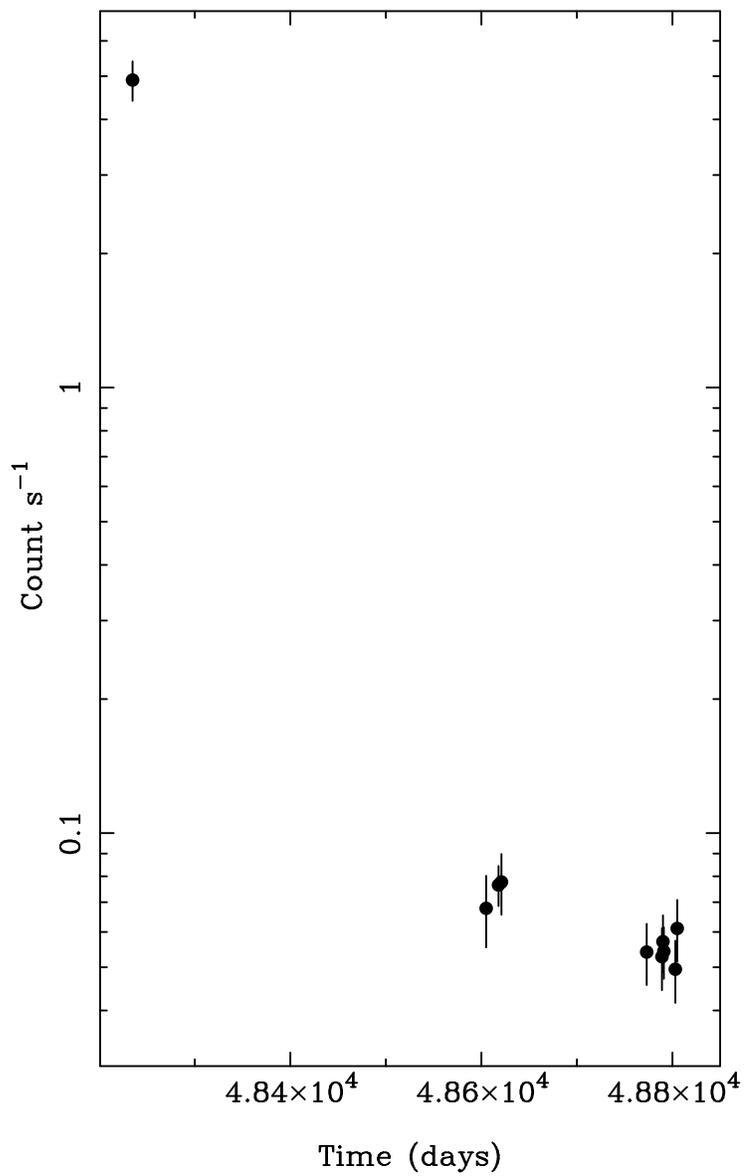

Fig. 3. Count rates during the all-sky survey and pointed *ROSAT* PSPC observations of RE J 1237+264. Note that the ordinate is logarithmic. From Brandt (1995).

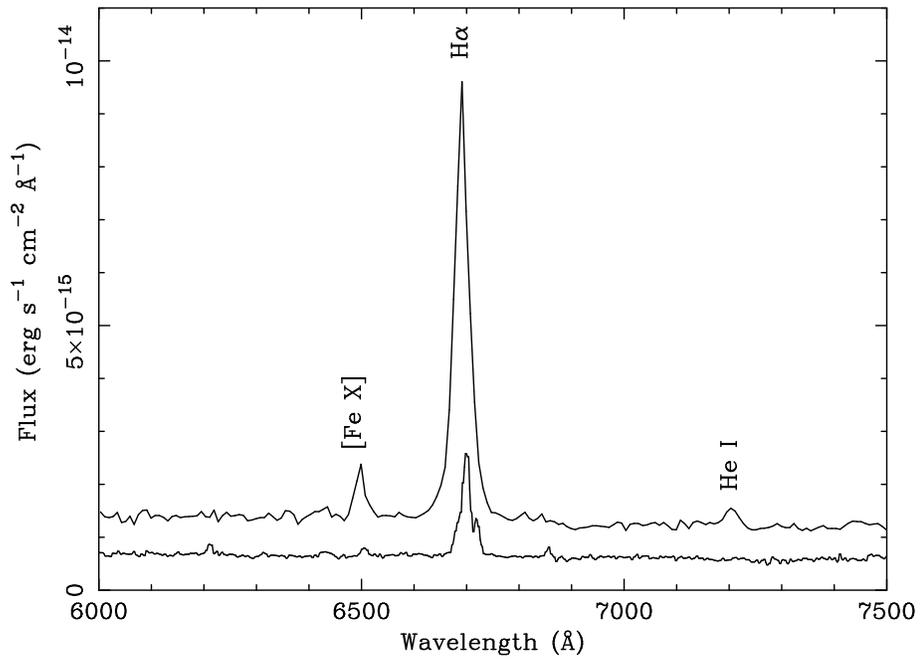

Fig. 4. Comparison of optical spectra of RE J 1237+264 taken in May 1991 (upper curve) and February 1992 (lower curve; from Bade et al. 1995). Note the 'decay' of the permitted lines after the soft X-ray outburst and the very strong high ionization iron line emission. From Brandt (1995).

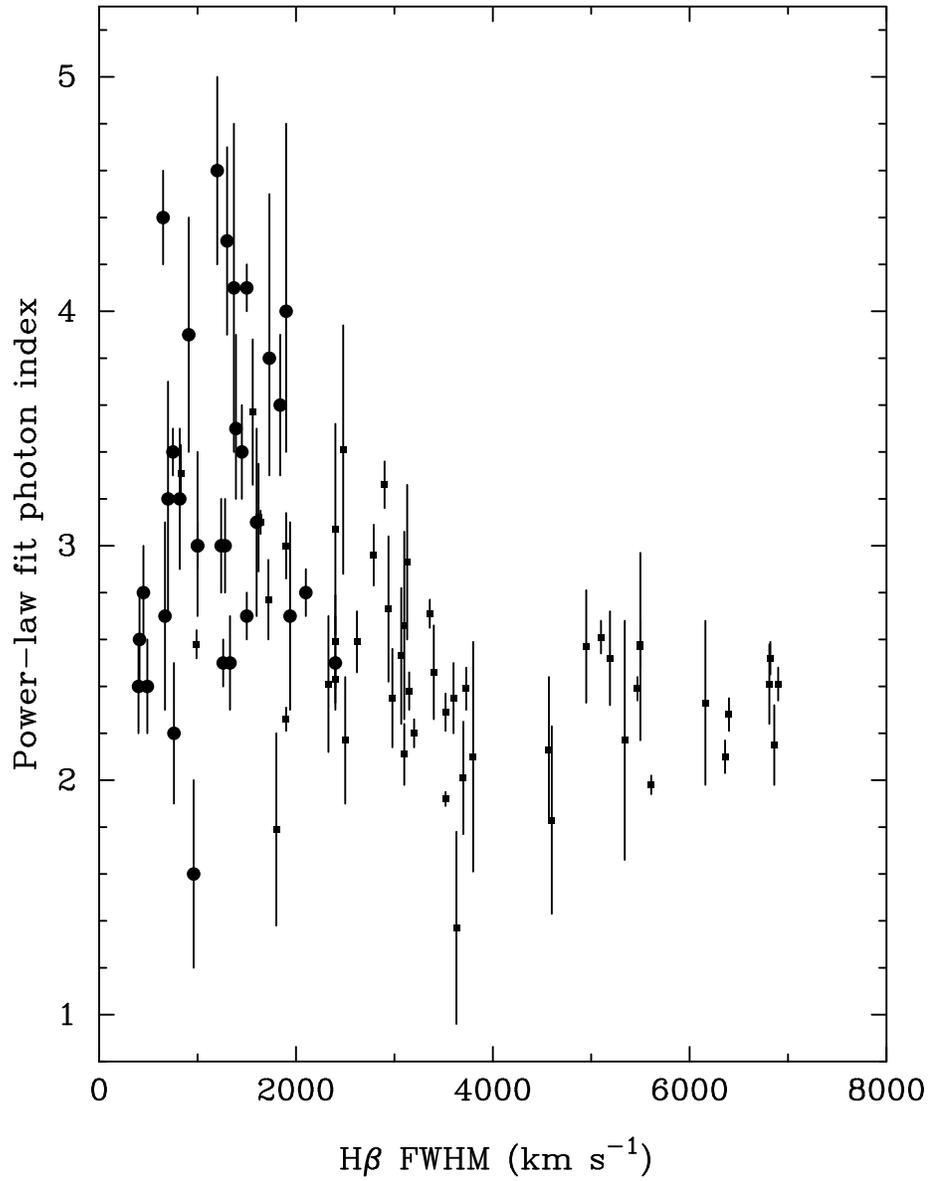

Fig. 5. *ROSAT* power-law fit photon index versus H$\beta$ FWHM for Seyfert 1 galaxies. Large solid dots denote Seyfert 1s that have been previously classified as NLS1 and small rectangles denote other Seyfert 1s. From BBF96 and Brandt (1995).